\documentclass[useAMS,usenatbib]{mn2e}

\usepackage{natbib}
\usepackage{graphicx}
\usepackage{amssymb}

\usepackage{color}

\title[Dusty outflows and CGM]{The connection between AGN-driven dusty outflows and the surrounding environment}

\author[ ]
{W. Ishibashi$^{1}$\thanks{E-mail:
wako.ishibashi@phys.ethz.ch} and A. C. Fabian$^{2}$
\footnotemark[0]\\
$^{1}$Institute for Astronomy, Department of Physics, ETH Zurich, Wolfgang-Pauli-Strasse 27, CH-8093 Zurich, Switzerland 
\footnotemark[0]\\
$^{2}$Institute of Astronomy, Madingley Road, Cambridge CB3 0HA 
}

\begin{document}

\pdfminorversion=4

\date{Accepted ? Received ?; in original form ? }

\pagerange{\pageref{firstpage}--\pageref{lastpage}} \pubyear{2012}

\maketitle

\label{firstpage}

\begin{abstract}
Significant reservoirs of cool gas are observed in the circumgalactic medium (CGM) surrounding galaxies. The CGM is also found to contain substantial amounts of metals and dust, which require some transport mechanism. We consider AGN (active galactic nucleus) feedback-driven outflows based on radiation pressure on dust. Dusty gas is ejected when the central luminosity exceeds the effective Eddington luminosity for dust. We obtain that a higher dust-to-gas ratio leads to a lower critical luminosity, implying that the more dusty gas is more easily expelled. Dusty outflows can reach large radii with a range of velocities (depending on the outflowing shell configuration and the ambient density distribution) and may account for the observed CGM gas. In our picture, dust is required in order to drive AGN feedback, and the preferential expulsion of dusty gas in the outflows may naturally explain the presence of dust in the CGM. On the other hand, the most powerful AGN outflow events can potentially drive gas out of the local galaxy group. We further discuss the effects of radiation pressure of the central AGN on satellite galaxies. AGN radiative feedback may therefore have a significant impact on the evolution of the whole surrounding environment. 
\end{abstract}

\begin{keywords}
black hole physics - galaxies: active - galaxies: evolution - galaxies: haloes - radiation: dynamics
\end{keywords}


\section{Introduction}

Recent observations indicate the presence of substantial amounts of cool gas in the circumgalactic medium (CGM) surrounding local galaxies and quasars at high redshift \citep{Thom_et_2012, Tumlinson_et_2013, Werk_et_2014, Prochaska_et_2013, Prochaska_et_2014}. 
In the local Universe, the COS-Halos survey is designed to study the CGM of $zÊ\sim 0.2$ galaxies by using background QSO sightlines \citep{Tumlinson_et_2013, Werk_et_2014}. 
At higher redshifts, the Quasar Probing Quasars (QPQ) survey, is dedicated to the study of the CGM around $z \sim 2$ quasars based on quasar pairs \citep{Prochaska_et_2014}.
These surveys reveal the common presence of cool gas in the CGM with high covering fractions and extending on scales of the virial radius.    

But the physical origin of the CGM itself is not clear, and the diffuse gas may arise from different sources, such as inflows from the intergalactic medium or outflowing galactic winds.  
Substantial amounts of metals and dust are also found in the CGM far beyond galaxies, on $\sim$hundred-kpc scales \citep{Peek_et_2014, Peeples_et_2014}. Heavy elements are by-products of stellar evolution and must have formed inside galaxies. The observed enrichment suggests that CGM gas previously cycled through galaxies, and some form of transport mechanism is required. 

Outflows provide a direct way of connecting the galaxies to the surrounding environment: galactic winds can eject interstellar gas and release metals in the haloes. 
Different forms of launching mechanisms, ranging from nuclear starbursts to active galactic nuclei (AGN), have been discussed in the literature. 
Galactic winds driven by momentum deposition have been studied in \citet{Murray_et_2005}; while \citet{Murray_et_2011} investigate the role of radiation pressure from star clusters in launching large-scale outflows of cool gas. 
Large-scale winds may also be driven by radiation pressure from self-gravitating discs \citep{Zhang_Thompson_2012}; and radiation pressure on dust grains has been considered as an important feedback mechanism in star-forming galaxies \citep{Andrews_Thompson_2011}.

In the context of AGN feedback, both kinetic- and radiative- modes are invoked to drive large-scale outflows via jets, winds, and radiation pressure \citep[][and references therein]{Fabian_2012}. 
Radio-mode feedback has strong observational evidence in terms of bubbles and cavities observed in galaxy clusters; on the other hand, the radiative-mode is less well documented due to obscuration effects. 
The importance of radiation pressure on dusty gas as a source of AGN feedback has been introduced by \citet{Fabian_1999}, and its role in driving galactic winds has been discussed e.g. in \citet{Murray_et_2005}. 
More recently, \citet{Thompson_et_2015} analyse the dynamics of dusty radiation pressure-driven shells, connecting with the observations of a variety of astrophysical sources. 
Focusing on AGN feedback, we have further studied the role of radiation pressure on dust in driving outflows on galactic scales \citep{Ishibashi_Fabian_2015}. In particular, we have shown that high-velocity outflows with large momentum ratios, comparable to the observed galactic-scale outflows, can be obtained by taking into account the effects of radiation trapping. 
Here we follow the subsequent evolution of the AGN radiation pressure-driven outflow on larger scales, and consider its impact on the surrounding environment. We briefly discuss how the propagation of dusty outflows and the presence of dust in the CGM may be naturally coupled within our framework. 


\section{The effective Eddington limit for dusty gas}
\label{Sect_Dusty_outflows}

We assume that radiation pressure on dust sweeps up the surrounding material into an outflowing shell, and consider a thin shell geometry following \citet{Thompson_et_2015}. 
The general form of the equation of motion of the shell is given by: 
\begin{equation}
\frac{d}{dt} (M_\mathrm{{sh}}(r)v) = \frac{L}{c} (1 + \tau_\mathrm{{IR}} - e^{-\tau_\mathrm{{UV}}}) - \frac{G M(r) M_\mathrm{{sh}}(r)}{r^2} 
\label{Eq_motion}
\end{equation} 
where $L$ is the central luminosity, $M(r)$ is the total mass distribution, and $M_\mathrm{{sh}}(r)$ is the shell mass.
The infrared (IR) and ultraviolet (UV) optical depths are respectively given by
\begin{equation}
\tau_\mathrm{{IR}}(r) = \frac{\kappa_\mathrm{{IR}} M_\mathrm{{sh}}(r)}{4 \pi r^2} 
\end{equation}
\begin{equation}
\tau_\mathrm{{UV}}(r) = \frac{\kappa_\mathrm{{UV}} M_\mathrm{{sh}}(r)}{4 \pi r^2} 
\end{equation} 
where $\kappa_\mathrm{{IR}}$ = 5 $\mathrm{cm^2 g^{-1}}$ $f_\mathrm{{dg, MW}}$ and $\kappa_\mathrm{{UV}}$ = $10^3 \mathrm{cm^2 g^{-1}}$ $f_\mathrm{{dg, MW}}$ are the IR and UV opacities, with the dust-to-gas ratio normalized to the Milky Way value. 
We note that the IR and UV opacities directly scale with the dust-to-gas ratio: $\kappa_\mathrm{{IR,UV}} \propto f_\mathrm{{dg}}$. 

A critical luminosity is defined by equating the outward force due to radiation pressure to the inward force due to gravity: 
\begin{equation}
L_\mathrm{E} = \frac{Gc}{r^2} M(r) M_\mathrm{{sh}}(r) (1 + \tau_\mathrm{{IR}} - e^{-\tau_\mathrm{{UV}}})^{-1}
\end{equation}
which can be considered as a generalised form of the effective Eddington luminosity. 

A major fraction of the AGN bolometric luminosity is emitted at ultraviolet (UV) wavelengths. The UV photons are absorbed by dust grains embedded in the gas, and subsequently re-emitted as infrared (IR) radiation. 
If the medium is optically thin to the reprocessed radiation, the IR photons freely escape without further interactions. 
The momentum flux transferred from the radiation field to the ambient gas is $L/c$ in the single scattering (SS) limit. 
In contrast, if the medium is optically thick to the reprocessed radiation, IR photons may be re-absorbed and re-emitted multiple times before diffusing out of the region. 
As a result of the radiation trapping, the photon energy density rises, with an associated increase in radiation pressure. 
Additional momentum will be deposited in the process, leading to boost factors exceeding the single scattering value ($\tau_{IR} L/c$, with $\tau_{IR} > 1$). 
The physics of the IR-optically thick regime has been previously investigated in different contexts \citep{Thompson_et_2005, Debuhr_et_2011, Roth_et_2012}. 

As discussed in e.g. \citet{Thompson_et_2015} and \citet{Ishibashi_Fabian_2015}, three main regimes can be distinguished according to the effective optical depth of the medium. 
If the shell is optically thick to the reprocessed IR radiation ($\tau_\mathrm{{IR}} > 1$), the corresponding Eddington luminosity is given by: 
\begin{equation}
L_\mathrm{{E,IR}} = \frac{4 \pi G c}{\kappa_\mathrm{{IR}}} M(r) 
\label{L_E_IR}
\end{equation} 
In the single scattering limit (i.e. optically thin to IR but optically thick to UV), the Eddington luminosity is given by: 
\begin{equation}
L_\mathrm{{E,SS}} = \frac{Gc}{r^2} M(r) M_\mathrm{{sh}}(r) 
\label{L_E_SS}
\end{equation} 
The effective Eddington luminosity in the regime where the shell becomes optically thin to UV radiation ($\tau_\mathrm{{UV}} < 1$) is: 
\begin{equation}
L_\mathrm{{E,UV}} = \frac{4 \pi G c}{\kappa_\mathrm{{UV}}} M(r) 
\label{L_E_UV}
\end{equation} 

Dusty gas is ejected when the central luminosity $L$ exceeds the effective Eddington luminosity in the appropriate limit. 
We see that in the IR-optically thick and UV-optically thin regimes, the effective Eddington luminosity is inversely proportional to the IR and UV opacity, respectively. Thus a higher opacity leads to a lower critical luminosity. 
In contrast, in the single scattering limit, the critical luminosity is independent of the medium opacity and hence of the dust-to-gas ratio. 
We note that $M_\mathrm{{sh}}(r)$ does not appear in Eq. (\ref{L_E_IR}) and Eq. (\ref{L_E_UV}), indicating that the shell mass configuration is irrelevant in these regimes.

\begin{figure}
\begin{center}
\includegraphics[angle=0,width=0.4\textwidth]{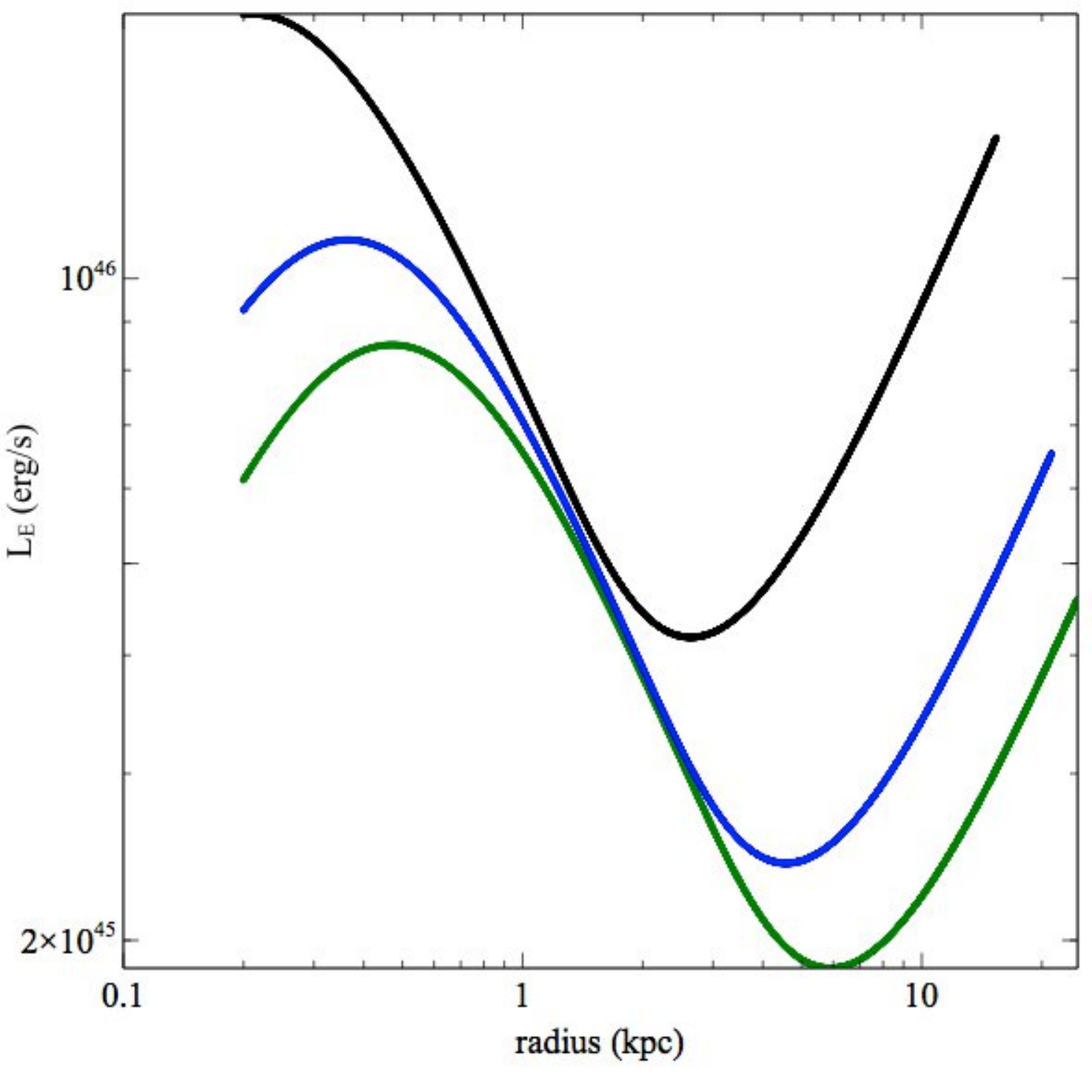} 
\caption{\small Effective Eddington luminosity as a function of radius for different values of the opacities in the case of the isothermal potential and fixed-mass shell. 
As fiducial parameters, we assume $\sigma$ = 200 km/s and $M_{sh} = 5 \times 10^{8} M_{\odot}$. 
Black curve: $\kappa_\mathrm{{IR}} = 5 \mathrm{cm^2 g^{-1}}$, $\kappa_\mathrm{{UV}} = 10^3 \mathrm{cm^2 g^{-1}}$ ($f_\mathrm{{dg}} = 1/150$, MW value), 
blue curve: $\kappa_\mathrm{{IR}} = 15 \mathrm{cm^2 g^{-1}}$, $\kappa_\mathrm{{UV}} = 3 \times 10^3 \mathrm{cm^2 g^{-1}}$ ($f_\mathrm{{dg}} = 1/50$),
green curve: $\kappa_\mathrm{{IR}} = 25 \mathrm{cm^2 g^{-1}}$, $\kappa_\mathrm{{UV}} = 5 \times 10^3 \mathrm{cm^2 g^{-1}}$ ($f_\mathrm{{dg}} = 1/30$)
}
\label{Fig_LE_r}
\end{center}
\end{figure}

As a specific example, we can consider the case of the isothermal potential ($M(r) = \frac{2 \sigma^2 r}{G}$, where $\sigma$ is the velocity dispersion) with a fixed-mass shell ($M_\mathrm{{sh}}(r) = M_\mathrm{{sh}}$). 
In this simple case, the three limits are respectively given by: \\
IR-optically thick regime ($\tau_\mathrm{{IR}} > 1$): 
\begin{equation}
L_\mathrm{{E,IR}} = \frac{8 \pi c \sigma^2 r}{\kappa_\mathrm{{IR}}} 
\end{equation} 
single scattering limit: 
\begin{equation}
L_\mathrm{{E,SS}} = \frac{2 \sigma^2 M_\mathrm{{sh}} c}{r} 
\label{L_E_SS_iso}
\end{equation} 
UV-optically thin regime ($\tau_\mathrm{{UV}} < 1$): 
\begin{equation}
L_\mathrm{{E,UV}} = \frac{8 \pi c \sigma^2 r}{\kappa_\mathrm{{UV}}} 
\end{equation}

Assuming the $M_\mathrm{{BH}} - \sigma$ relation given in \citet{Kormendy_Ho_2013} and taking $\sigma = 200$km/s as a fiducial value, the standard Eddington luminosity is $L_\mathrm{E} = \frac{4 \pi G c m_p}{\sigma_T} M_\mathrm{{BH}} \sim 4 \times 10^{46}$erg/s.
In Figure \ref{Fig_LE_r} we plot the effective Eddington luminosity as a function of radius for different values of the IR and UV opacities. 
We see that enhanced opacities (due e.g. to higher dust-to-gas ratios) lead to lower critical luminosities, as expected. We also observe that the three curves roughly overlap at intermediate radii on $\sim$kpc-scales. This region in fact corresponds to the single scattering regime, where the critical luminosity is independent of the medium opacity, as seen from Eq. (\ref{L_E_SS_iso}). 
Since the IR and UV opacities directly scale with the dust-to-gas ratio, an increase in $f_\mathrm{{dg}}$ by a given factor implies a decrease in the critical luminosity by the corresponding factor. 
Thus a medium having a higher dust-to-gas fraction is preferentially ejected, as it requires only a lower luminosity. 
This implies that, for a given central luminosity, the more dusty material will be more easily expelled from the galaxy. 
This characteristic behaviour may have significant implications for the presence of dust observed in the CGM around galaxies (see Discussion). 

We have previously analysed the dependence of the shell dynamics on the underlying physical parameters \citep{Ishibashi_Fabian_2015}. 
We recall that enhanced opacities lead to higher shell velocities, with the shell asymptotic velocity scaling as $v_{\infty} \propto \kappa_\mathrm{{UV}}^{1/4}$. 
Thus dusty shells can be accelerated to higher speeds and reach greater distances, spreading dust on larger scales.


\section{Propagation of dusty outflows on large scales}
\label{Sect_Propagation}

We are now interested in the subsequent evolution of the outflowing shell and its impact on the large-scale environment. 
The gravitational potential is modelled by a NFW density profile: 
\begin{equation}
\rho(r) = \frac{\rho_\mathrm{c} \delta_\mathrm{c}}{(r/r_\mathrm{s})(1+r/r_\mathrm{s})^2} \, , 
\end{equation} 
where $\rho_\mathrm{c} = \frac{3 H^2}{8 \pi G}$ is the critical density, $\delta_\mathrm{c}$ is a characteristic density, and $r_\mathrm{s}$ is a scale radius. 
The corresponding mass profile is given by 
\begin{equation}
M(r) = 4 \pi \delta_\mathrm{c} \rho_\mathrm{c} r_\mathrm{s}^3 \, \left[ \ln \left(1+\frac{r}{r_\mathrm{s}}\right) - \frac{r}{r+r_\mathrm{s}} \right] \, .
\end{equation} 


\subsection{Fixed-mass shells}

\begin{figure}
\begin{center}
\includegraphics[angle=0,width=0.4\textwidth]{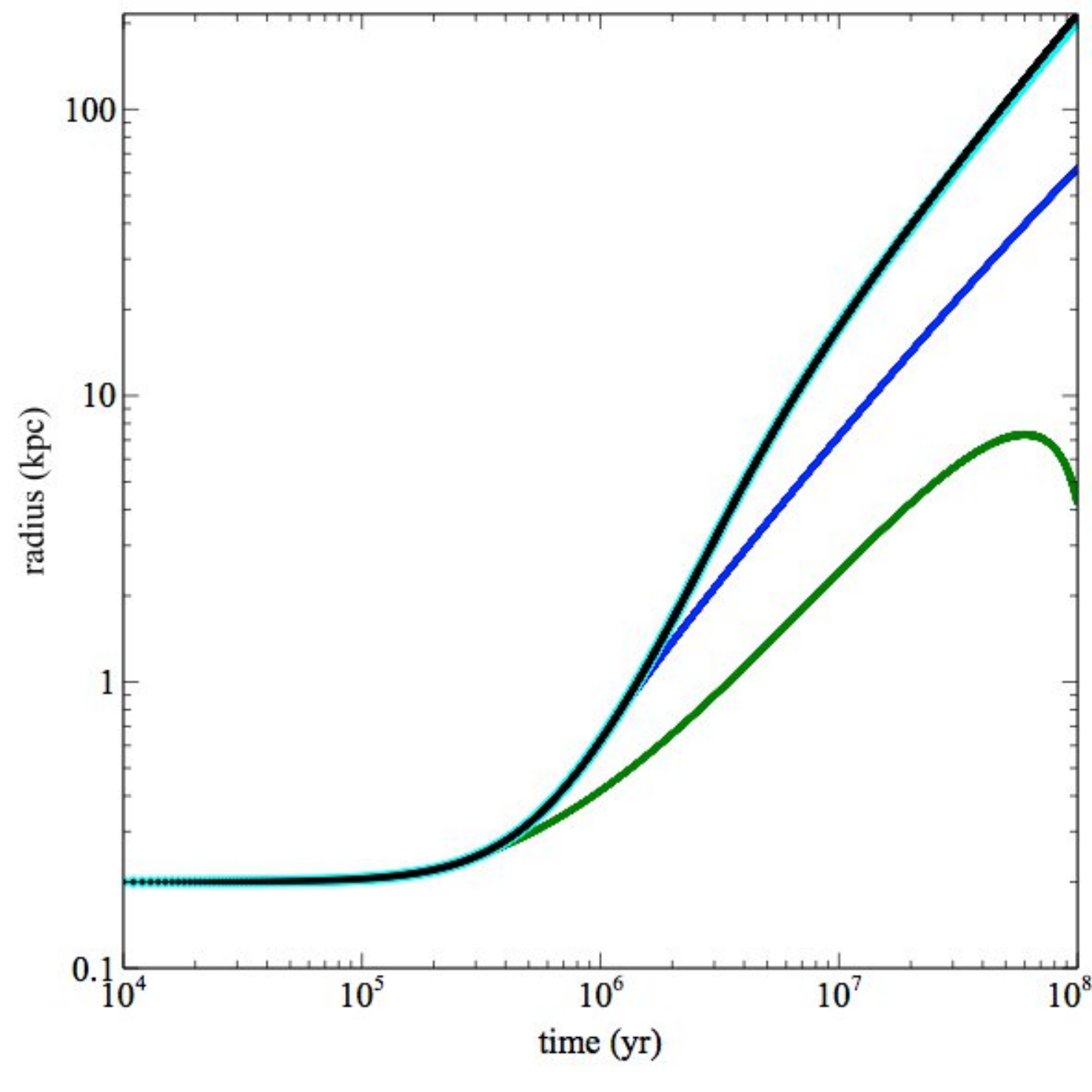} 
\caption{\small 
Radius as a function of time for different values of the central AGN lifetime. 
$\Delta t_\mathrm{{AGN}} = 10^8$yr (black), 
$\Delta t_\mathrm{{AGN}} = 10^7$yr (cyan), 
$\Delta t_\mathrm{{AGN}} = 10^6$yr (blue), 
$\Delta t_\mathrm{{AGN}} = 2.5 \times 10^5$yr (green). 
}
\label{Fig_r_t_varTime}
\end{center}
\end{figure}
\begin{figure}
\begin{center}
\includegraphics[angle=0,width=0.4\textwidth]{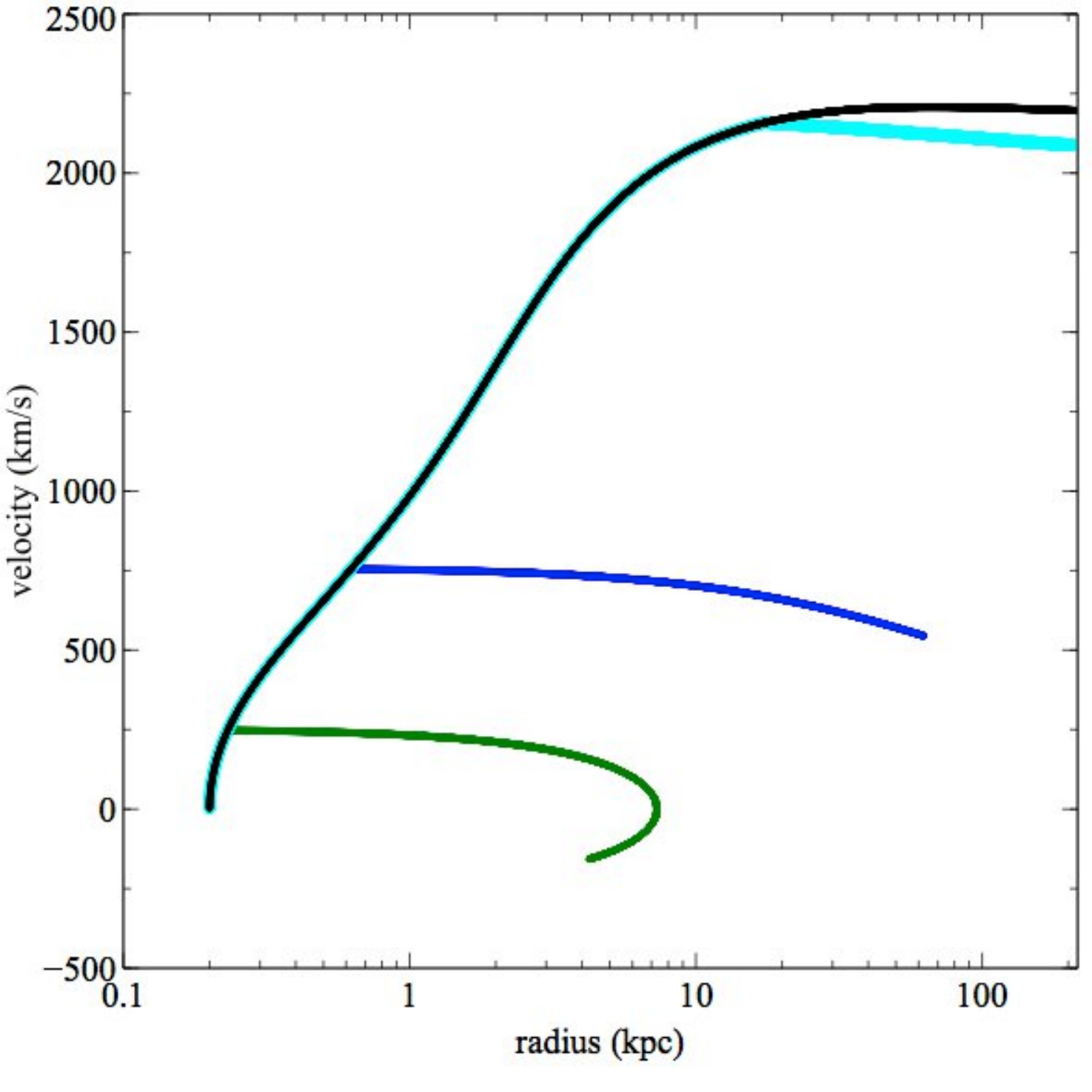} 
\caption{\small 
Velocity as a function of radius for different values of the central AGN lifetime. 
$\Delta t_\mathrm{{AGN}} = 10^8$yr (black),  
$\Delta t_\mathrm{{AGN}} = 10^7$yr (cyan), 
$\Delta t_\mathrm{{AGN}} = 10^6$yr (blue), 
$\Delta t_\mathrm{{AGN}} = 2.5 \times 10^5$yr (green). 
}
\label{Fig_v_r_varTime}
\end{center}
\end{figure}

We first consider the case of a fixed-mass shell. The fiducial parameters are $L = 5 \times 10^{46}$erg/s, $M_\mathrm{{sh}} = 5 \times 10^{8} M_{\odot}$, $\kappa_\mathrm{{IR}}$ = 5 $\mathrm{cm^2/g}$, $\kappa_\mathrm{{UV}}$ = $10^3$ $\mathrm{cm^2/g}$. 
In Fig. \ref{Fig_r_t_varTime}, we plot the temporal evolution of the shell for different values of the central AGN lifetime, $\Delta t_\mathrm{{AGN}}$, and Fig. \ref{Fig_v_r_varTime} shows the corresponding radial velocity profiles. 
We see that if the active phase lasts for $10^8$yr (black curve), the shell reaches large radii exceeding $\gtrsim 200$kpc with high velocity ($> 2000$km/s). 
Even if the central AGN switches off after $10^7$yr (cyan curve), the shell still carries on, and reaches fairly large radii ($\sim 200$kpc) with high speed ($\gtrsim 2000$km/s).
Fixed-mass shells can thus reach $\sim$few hundred kiloparsec-scale radii within typical AGN activity timescales, comparable to the Salpeter time ($t_\mathrm{s} \sim 5 \times 10^7$yr). 
These powerful outflow events can potentially escape the host halo and affect the surrounding environment on larger scales. 
For a shorter AGN phase ($\Delta t_\mathrm{{AGN}} = 10^6$yr, blue curve), the shell travels at a lower speed ($\sim 500$km/s) and extends to smaller radii ($\sim 60$kpc). 
Such outflowing gas may be associated with CGM gas observed at intermediate radii with modest velocities. 
If the AGN activity timescale is further reduced, the shell reaches a certain maximal distance and then falls back (green curve). The shell thus remains trapped in the halo and may be considered a form of galactic fountain. 
Therefore a wide range of shell behaviours can be obtained just by considering variations in the central AGN lifetime.

\begin{figure}
\begin{center}
\includegraphics[angle=0,width=0.4\textwidth]{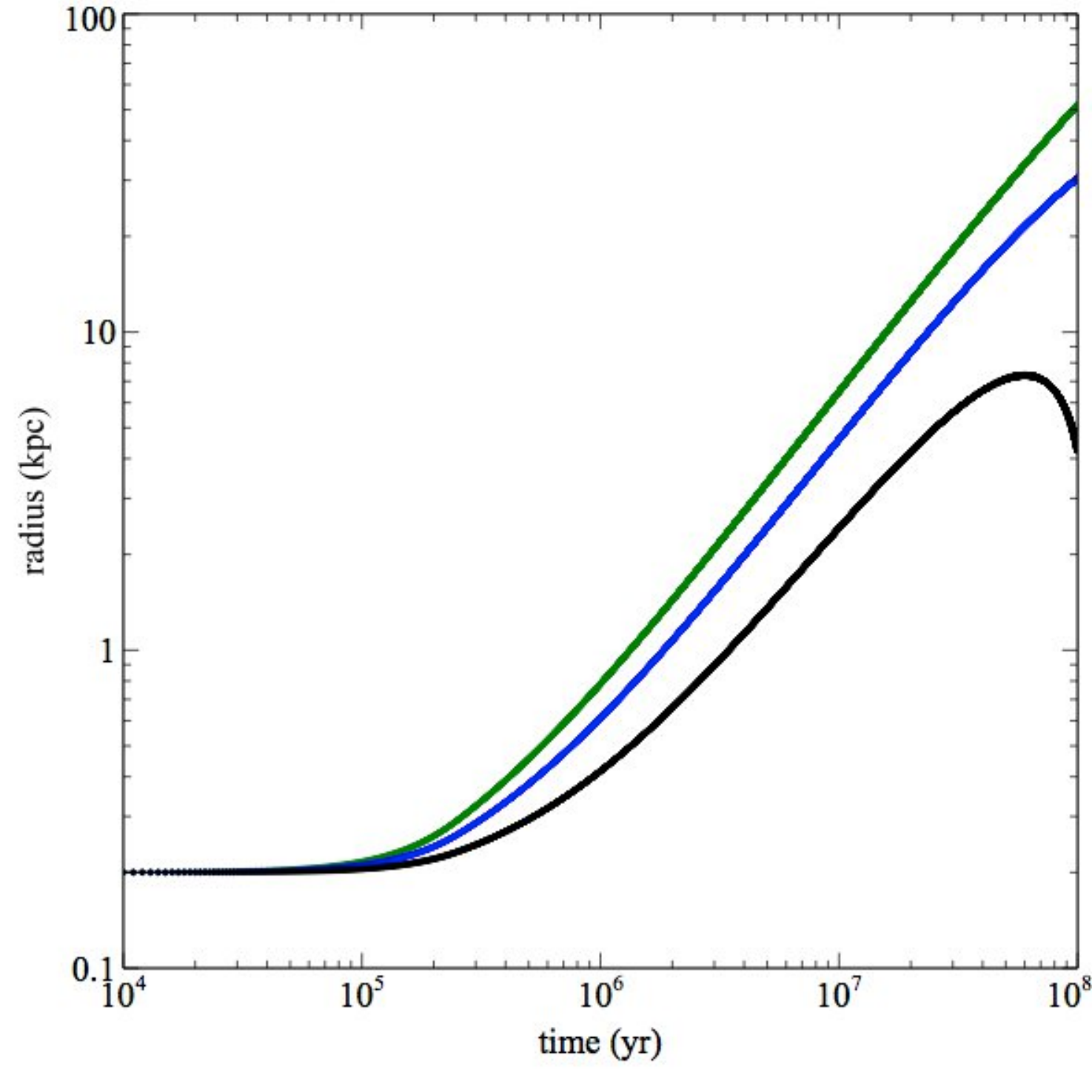} 
\caption{\small 
Radius as function of time in the case of $\Delta t_\mathrm{{AGN}} = 2.5 \times 10^5$yr and variations of the IR and UV opacities. 
$\kappa_\mathrm{{IR}} = 5 \mathrm{cm^2 g^{-1}}$, $\kappa_\mathrm{{UV}} = 10^3 \mathrm{cm^2 g^{-1}}$ (black).
$\kappa_\mathrm{{IR}} = 15 \mathrm{cm^2 g^{-1}}$, $\kappa_\mathrm{{UV}} = 3 \times 10^3 \mathrm{cm^2 g^{-1}}$ (blue).
$\kappa_\mathrm{{IR}} = 25 \mathrm{cm^2 g^{-1}}$, $\kappa_\mathrm{{UV}} = 5 \times 10^3 \mathrm{cm^2 g^{-1}}$ (green). 
}
\label{Fig_r_t_varKappa}
\end{center}
\end{figure}

As we have seen in Section \ref{Sect_Dusty_outflows}, the effective Eddington luminosity is determined by the dust content, and thus variations in the local medium opacity can significantly modify the shell dynamics. 
In Fig. \ref{Fig_r_t_varKappa}, we plot the temporal evolution of the shell for different values of the IR and UV opacities (or equivalently dust-to-gas ratios). 
For enhanced opacities, the critical luminosity is reduced and the shell escape is facilitated. 
In fact, we see that the shell may continue to flow outwards instead of falling back, simply by increasing the dust opacities.
The radial extent of the outflow also depends on the dust content, implying that the more dusty gas can be transported to larger radii. 
Actually, the widespread presence of dust on $\sim$10kpc-scales may often be observed after the central AGN has switched off.


\subsection{Expanding shells}

\begin{figure}
\begin{center}
\includegraphics[angle=0,width=0.4\textwidth]{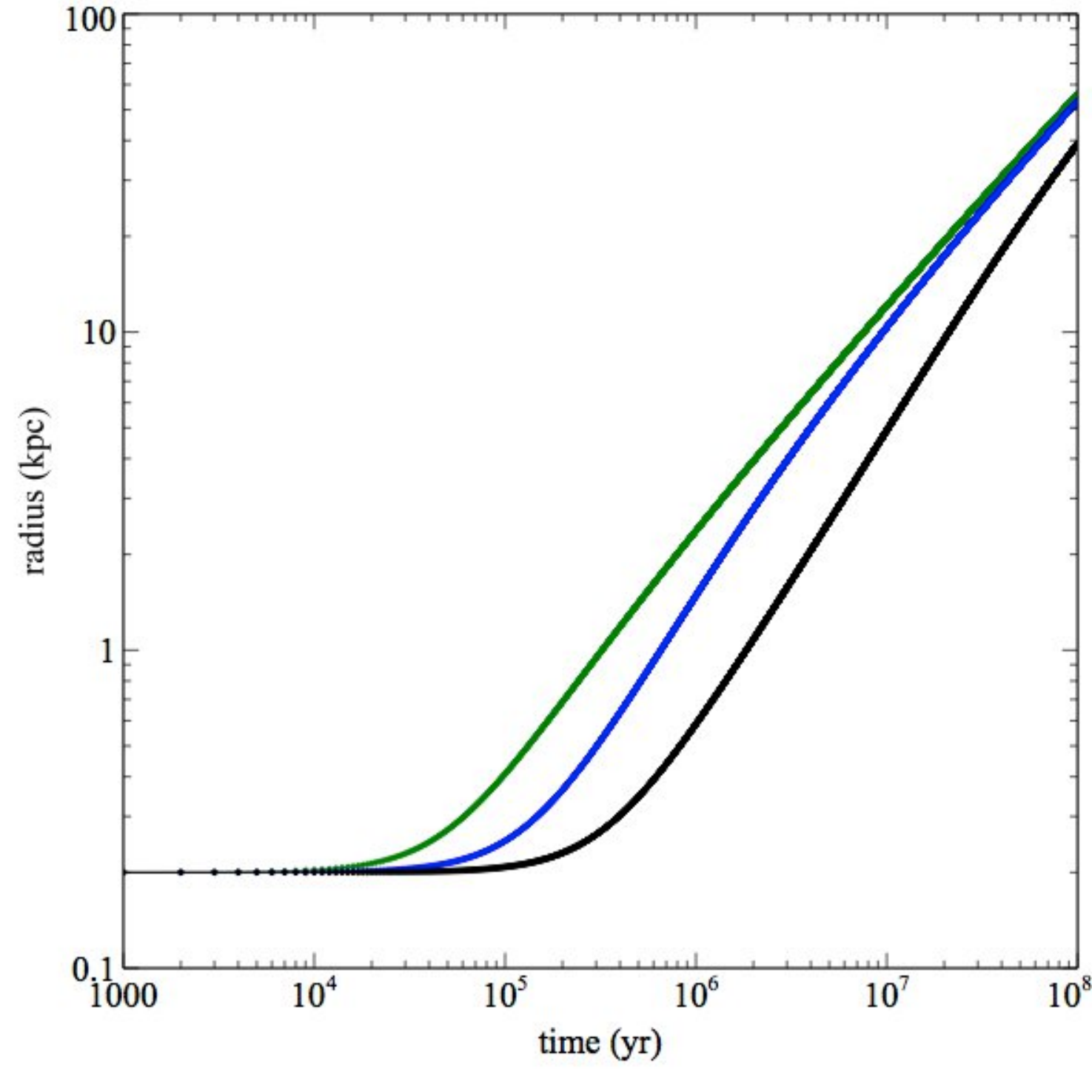} 
\caption{\small 
Radius as a function of time of a shell sweeping up matter from the surrounding environment for different values of the external density ($\Delta t_\mathrm{{AGN}} = 10^8$yr). 
$n_0 = 100 \mathrm{cm^{-3}}$ (black), 
$n_0 = 10 \mathrm{cm^{-3}}$ (blue), 
$n_0 = 1 \mathrm{cm^{-3}}$ (green).  
}
\label{Fig_plot_r_t_varN0}
\end{center}
\end{figure}
\begin{figure}
\begin{center}
\includegraphics[angle=0,width=0.4\textwidth]{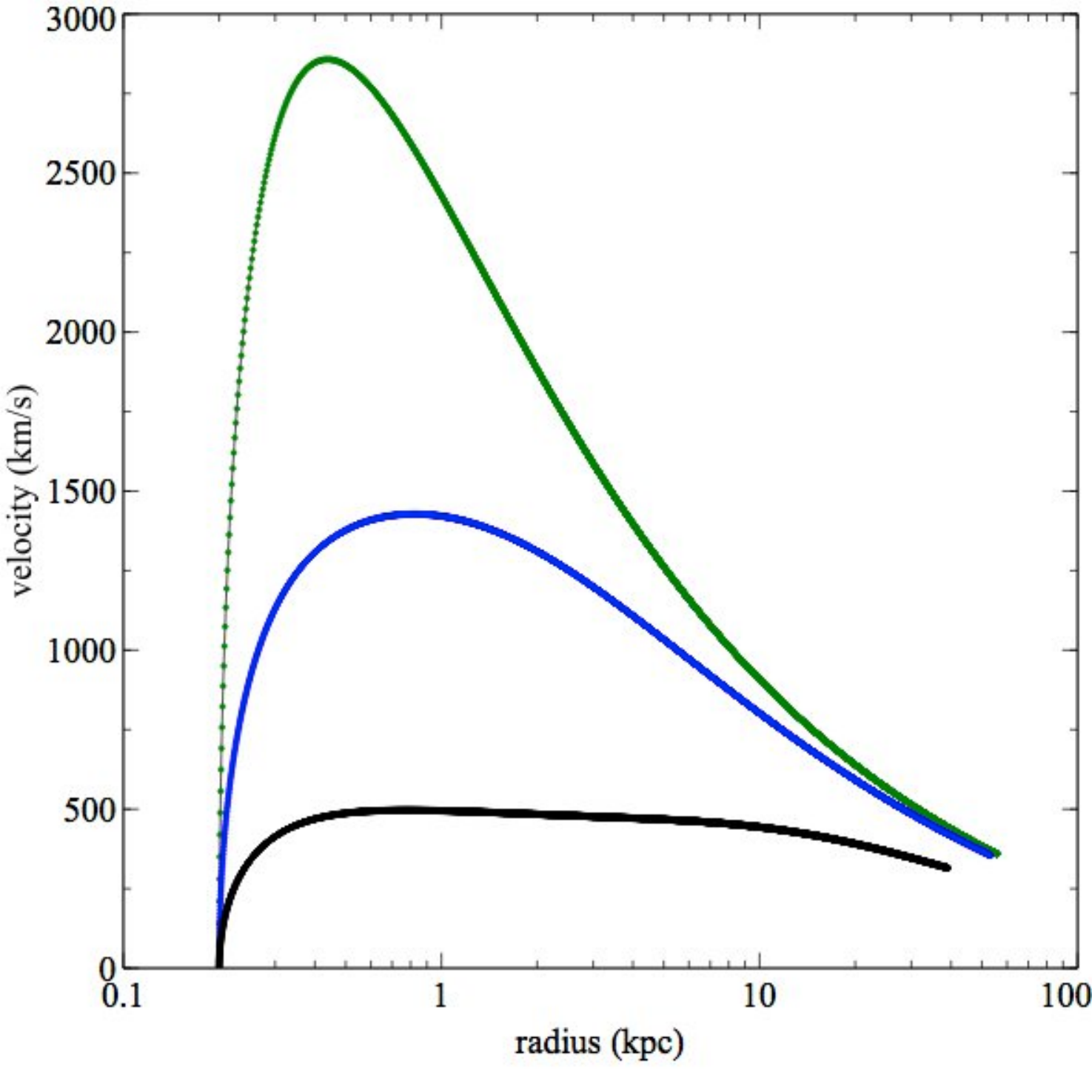} 
\caption{\small 
Velocity as a function of radius of a shell sweeping up matter from the surrounding environment for different values of the external density ($\Delta t_\mathrm{{AGN}} = 10^8$yr). 
$n_0 = 100 \mathrm{cm^{-3}}$ (black), 
$n_0 = 10 \mathrm{cm^{-3}}$ (blue), 
$n_0 = 1 \mathrm{cm^{-3}}$ (green).
}
\label{Fig_plot_v_r_varN0}
\end{center}
\end{figure}

We next consider a shell sweeping up matter from the surrounding environment. 
The density distribution of the ambient medium can be parametrized as a power law of radius with slope $\alpha$:
\begin{equation}
n(r) = n_0 \left( \frac{r}{R_0} \right)^{-\alpha}
\end{equation}
where $n_0$ is the density of the external medium.
The corresponding swept-up mass is given by:
\begin{equation}
M_\mathrm{{sw}}(r) = 4 \pi m_p \int n(r) r^2 dr = 4 \pi m_p n_0 R_0^{\alpha} \frac{r^{3-\alpha}}{3-\alpha}
\end{equation}
We consider here an isothermal distribution, with $\alpha = 2$, for which the swept-up mass scales with radius as: 
\begin{equation}
M_\mathrm{{sh}}(r) = 4 \pi m_p R_0^2 n_0 r   \propto r
\end{equation}

In Fig. \ref{Fig_plot_r_t_varN0}, we plot the temporal evolution of the expanding shells for different values of the external density. The corresponding radial velocity profiles are shown in Fig. \ref{Fig_plot_v_r_varN0}. 
As the expanding shells sweep up matter, they are slowed down and propagate to smaller radii compared to the case of fixed-mass shells. If the density of the ambient medium is lowered, the shells can reach higher velocities and attain somewhat larger radii. 
However, this trend only holds up to a certain point. In fact, below a certain critical density, the shell enters in the optically thin regime. In this limit, the driving term is dominated by the UV term: 
\begin{equation}
a_\mathrm{{UV}} \sim \frac{\kappa_\mathrm{{UV}} L}{4 \pi c r^2} 
\end{equation}
which is independent of $n_0$. 
Thus a further reduction of the external density has no effect on the shell dynamics. 
The expanding shells typically reach radial distances on the order of $\sim 50$kpc with velocities of $\lesssim 500$km/s. 
The outflowing gas may thus be associated with the CGM gas observed on similar spatial scales with modest speeds. 

As in the case of fixed-mass shells, enhanced opacities allow the expanding shells to reach somewhat higher velocities and hence greater distances. But, as the expanding shells continue to sweep up matter, they are decelerated and likely to remain bound on larger scales.


\section{The effects of AGN radiation pressure on satellite galaxies}
\label{Sect_Satellites}

Radiation pressure from the central AGN may also affect surrounding satellite galaxies. 
Consider a central source with black hole mass $M$ and corresponding Eddington luminosity $L_\mathrm{E}$; and a satellite with mass $m$ and radius $d$, located at a distance $D$ from the central host.
The balance between the radiative force due to the central source and the satellite gravitational force is given by
\begin{equation}
\frac{L_\mathrm{e} \sigma_\mathrm{e}}{4 \pi D^2 c} = \frac{G m m_p}{d^2}
\end{equation}
where $\sigma_e$ is the effective interaction cross section and $L_\mathrm{e}$ the equivalent Eddington luminosity of the satellite galaxy. 
Introducing the standard Eddington limit for the central source, $L_\mathrm{E} = \frac{4 \pi G c m_p}{\sigma_\mathrm{T}} M$ (where $\sigma_\mathrm{T}$ is the Thomson cross section), we can write:
\begin{equation}
\frac{L_\mathrm{e}}{L_\mathrm{E}} = (\frac{\sigma_T}{\sigma_\mathrm{e}}) (\frac{m}{M}) (\frac{D}{d})^2 
\label{Eq_LE_ratio}
\end{equation}
If $L_\mathrm{e}/L_\mathrm{E} \leq 1$, radiation pressure from the central source may remove gas from the satellite galaxy. 
For $D/d \sim 100$ and taking the Thomson cross section ($\sigma_\mathrm{e} \sim \sigma_\mathrm{T}$), the required mass ratio is $m/M \sim 10^{-4}$. 
On the other hand, assuming a typical dust absorption cross section ($\sigma_\mathrm{e} \sim \sigma_\mathrm{d} \sim 10^3 \sigma_\mathrm{T}$), the corresponding mass ratio is $m/M \sim 10^{-1}$. 
In fact, we see from Eq. (\ref{Eq_LE_ratio}) that the action of radiation pressure on dust introduces a factor of $\sigma_\mathrm{d}/\sigma_\mathrm{T} \sim 10^3$ in the coupling enhancement. 

Considering typical values for the satellite galaxy size ($d \sim 300$pc) and distance to the the central host ($D \sim 100$kpc) as observed in local groups \citep[][and references therein]{McConnachie_2012, Collins_et_2013, Belokurov_2013}, the corresponding mass ratio would be $m/M \sim 10^{-2}$. Thus a central host of mass $M \sim 10^8 M_{\odot}$ can potentially eject dusty gas from a satellite of mass $m \sim 10^6 M_{\odot}$ located at a distance of $\sim 100$kpc. 
Dusty gas may be more easily disrupted from extended systems, while it may be retained for more massive, compact configurations. 

In addition, if the quasar emission follows a bipolar pattern, as expected if radiation originates from an accretion disc, then satellite galaxies orbiting in the equatorial plane of that pattern may escape gas stripping. This might account for the anisotropic distribution of satellites observed around a number of local galaxies (see Discussion).


\section{Discussion}

\subsection{The CGM gas reservoirs}

Substantial amounts of cool gas are reported in recent observations of the CGM surrounding galaxies and quasars \citep{Tumlinson_et_2013, Werk_et_2014, Prochaska_et_2013, Prochaska_et_2014}. 
Strong neutral hydrogen is detected in local galaxies in the COS-Halos survey. The bulk of the cool gas, located within impact parameters of $\sim 150$kpc, is likely to be gravitationally bound to the galaxies \citep{Tumlinson_et_2013}. 
At higher redshifts, massive reservoirs of cool and metal-enriched gas are detected in $z \sim 2$ quasar host haloes.
The CGM is observed to extend to the virial radius ($R_V \sim 160$kpc); and the typical velocities are on the order of a few hundred km/s \citep{Prochaska_et_2013}. 
Based on the QPQ survey, \citet{Prochaska_et_2014} suggest that the cool CGM surrounding $z \sim 2$ quasars may be the `pinnacle' among all galaxy populations. 

Cosmological simulations, including stellar feedback, indicate that cool gas can be directly ejected into the haloes by galactic winds \citep{Faucher-Giguere_et_2015}. The resulting covering fractions of neutral hydrogen are consistent with the values observed around Lyman break galaxies at $z \sim 2$. But the predicted values cannot account for the high covering fractions measured around $z \sim 2$ quasars in massive haloes \citep{Faucher-Giguere_et_2015}.
This suggests that the large amount of cool gas observed in quasar host haloes is somehow linked to the presence of the central AGN. However, it has been argued that quasar-driven outflows can not reach $\gtrsim 100$-kpc scales within typical activity timescales, and therefore AGN only play a minor role in explaining the observed CGM \citep{Prochaska_et_2014}.

In Section \ref{Sect_Propagation}, we have seen that radiation pressure-driven outflows can reach large radii within plausible AGN activity timescales. 
In particular, fixed-mass shells can travel to radii comparable to the virial radius on timescales of the order of the Salpeter time. 
High-velocity shells can potentially escape the host halo and propagate into the intergalactic medium. 
In contrast, expanding shells sweeping up ambient material are considerably slowed down, and typically reach radii of $\sim 50$kpc, with a range of velocities depending on the density of the external medium.  
This outflowing material may be associated with the observed CGM gas moving at moderate velocities. 
Dusty outflows may also contribute to the enrichment of the CGM, with the dust content influencing the physical extent of the outflow itself. 


\subsection{Dust and metals in the CGM}

The CGM surrounding galaxies can be a substantial repository of metals. 
Results based on the COS-Halos survey indicate that the bulk of metals reside outside of the galaxies, with a significant fraction of CGM metals found in the solid phase, in the form of CGM dust \citep{Peeples_et_2014}. 
Based on SDSS data, \citet{Peek_et_2014} detect an excess reddening due to the presence of dust in the CGM of low-redshift galaxies. The reddening signal is detected on scales of $\sim 150$kpc and is observed to fall off at larger radii. 
Large amounts of dust are thus distributed in a large-scale configuration around nearby galaxies. 

The widespread detection of metals in the CGM, which are presumably produced during stellar evolution, requires an efficient transport mechanism. 
Since the metals are mostly in the form of dust particles, and it is the dust itself which experiences the radiation force, they can be naturally propagated into the CGM through radiation pressure-driven outflows. 
As the effective Eddington luminosity scales inversely with the dust opacity, the critical luminosity for expulsion decreases with increasing dust-to-gas ratio. Thus a medium with a higher dust content is preferentially ejected. 
This accounts for the presence of dust in the CGM surrounding galaxies in a self-consistent way. 
Moreover, enhanced opacities also lead to higher shell velocities, which can transport dust to farther distances. 
In our picture, the large-scale distribution of dust in the CGM is a natural consequence of dusty outflows, as the underlying driving mechanism is based on radiation pressure on dust. 
This in turn requires a sufficient amount of dust in order to drive efficient AGN feedback. 

Recent observational findings indicate that large quantities of dust can be produced in core-collapse supernovae. 
High dust masses are estimated in the supernova remnants SN 1987A \citep{Wesson_et_2015} and the Crab Nebula \citep{Owen_Barlow_2015}. The implied dust-to-gas mass ratio is of the order of $1/30$ in the case of the Crab Nebula. 
Additional processes, such as grain growth in the interstellar medium and coagulation of pre-existing dust grains, may also contribute to the dust mass increase at late times \citep{Michalowski_2015, Wesson_et_2015}. 
Moreover, fits to the observed spectral energy distributions favour large dust grains, which are more likely to resist destruction by sputtering, suggesting that a significant fraction of the dust formed in supernova ejecta can actually survive \citep{Owen_Barlow_2015}. 
We have previously discussed the possibility of star formation triggered in AGN radiation pressure-driven outflows \citep{Ishibashi_Fabian_2012}. In such a scenario, fresh dust can be released when massive stars (formed within the outflowing shell) explode as supernovae. 
The additional dust spread and mixed into the surrounding environment may contribute to enhance the overall feedback process and further sustain the propagation of dusty outflows.

\subsection{Impact of AGN radiative feedback on the group environment}

We have seen that AGN feedback can also affect other galaxies residing in the surrounding environment, and in particular satellite galaxies orbiting the central host. 
As discussed in Section \ref{Sect_Satellites}, radiation pressure from a central object of mass $M \sim 10^8 M_{\odot}$ is potentially able to remove dusty gas from a dwarf satellite with mass $m \sim 10^6 M_{\odot}$ located at a distance of $\sim 100$kpc. 
As an example, such a scenario may be applied to the case of the Andromeda galaxy (M31) and its dwarf satellites. 
In fact, M31 is believed to harbour a massive object at its centre, with an estimated black hole mass of the order of $\sim 10^8 M_{\odot}$, which presents strong X-ray flux variations similar to the flaring X-ray emission behaviour of Sgr A* in the Galactic Center \citep{Li_et_2011}. The lopsided stellar disc observed in the central $\sim$pc-region of M31 has been interpreted as the stellar remnant of eccentric discs that are often found in numerical simulations \citep{Hopkins_Quataert_2010}. Such eccentric discs, formed by gravitational instabilities, are thought to be responsible for the efficient transfer of angular momentum which allowed quasar-like accretion rates in the past. 

As mentioned in Section \ref{Sect_Satellites}, if AGN feedback follows a bipolar configuration, the resulting influence on the satellite galaxies is different depending on their relative location with respect to the central AGN. For instance, dwarf satellites orbiting in the equatorial plane of the inner disc or torus may be shielded and thus able to retain their gas, contrary to other satellites that are stripped of most of their gas. This may account for the planar structures of dwarf satellites observed around Andromeda and possibly other galaxies \citep[][and references therein]{Ibata_et_2013, Pawlowski_et_2014}. Therefore AGN feedback may even influence the spatial distribution of surrounding satellite galaxies, and the long-term imprints left by such feedback effects could be observed in the Local Group. 

Strong AGN feedback may also affect the gas content of the local group environment by driving dusty gas out. 
In the most extreme cases, much of the intragroup gas may be ejected by some powerful outflow event; and this might partly account for the deficiency of gas and the low baryon fraction observed in galaxy groups. 
The effects of a central AGN should in fact be more pronounced in galaxy groups, which have shallower potential wells, compared to galaxy clusters. 
Indeed, the baryon fraction is observed to be significantly lower than the cosmic value in galaxy groups, unlike in clusters of galaxies \citep[e.g.][]{Giodini_et_2010}. 

In the case of galaxy clusters, the impact of kinetic-mode feedback is clearly visible as e.g. giant radio lobes extending over several hundred kiloparsecs and affecting the whole intracluster medium \citep[][]{Fabian_2012, McNamara_Nulsen_2012}. 
In contrast, the role of radiative-mode feedback and its influence on the large-scale environment is much more elusive. 
We have previously discussed how radiation pressure on dust may shape the global properties of galaxies, setting the characteristic radius and mass \citep{Ishibashi_Fabian_2014}. 
In Section \ref{Sect_Propagation}, we have seen that radiation pressure-driven outflows can propagate to larger radii, suggesting that the presence of a central AGN can affect gas on hundred kpc scales.
Radiative feedback may thus be equally important in shaping the large-scale environment of galaxies, without requiring relativistic jets as in radio-loud objects (which only form a minority of the AGN population) and may have a wider applicability. 
Feedback from the central AGN may therefore affect, not only the evolution of its own host galaxy, but also the development of satellite galaxies and the surrounding group environment.

\section*{Acknowledgements}

WI acknowledges support from the Swiss National Science Foundation.   
ACF acknowledges ERC Advanced Grant FEEDBACK.

\bibliographystyle{mn2e}
\bibliography{biblio.bib}

\label{lastpage}

\end{document}